\documentclass[11pt]{article}

\usepackage[T1]{fontenc}
\usepackage[utf8]{inputenc}
\usepackage[english]{babel}
\usepackage{amsmath,amssymb,amsfonts,amsthm}            
\usepackage{mathtools,mathrsfs}                         
\usepackage{thmtools,tensor,dsfont,bm}                  
\usepackage{geometry,fancyhdr}                          
\usepackage{graphicx,subcaption}                        
\usepackage[labelfont=small]{caption}                   
\usepackage[makeroom]{cancel}                           
\usepackage[dvipsnames,svgnames,x11names,table]{xcolor} 
\usepackage{cite}                                       
\usepackage[pdfencoding=auto, psdextra]{hyperref}       
\usepackage[capitalize]{cleveref}                       

\newcommand{\sfrac}[2]{{\textstyle \frac{#1}{#2}}}      
\newcommand{\gi}[1]{\text{\tiny \rm #1}}                
\newcommand{\Conf}{\text{Conf}}                         
\newcommand{\conf}{\mathfrak {conf}}                    
\newcommand{\M}{\mathds{M}}                            
\newcommand{\D}{\mathds{D}}                            
\newcommand{\T}{\mathds{T}}                            
\newcommand{\K}{\mathds{K}}                            
\newcommand{\X}{\mathds{X}}                            
\newcommand{\I}{\mathds{I}}                            

\usepackage{xcolor, soul}
\sethlcolor{red}

\newenvironment{psm}{\left(\begin{smallmatrix}}{\end{smallmatrix}\right)}

\newcommand\myshade{90}
\colorlet{mycitecolor}{JungleGreen}
\colorlet{mylinkcolor}{purple}
\colorlet{myurlcolor}{MediumSeaGreen}
\hypersetup{
    colorlinks = true,
    citecolor  = mycitecolor!\myshade!black,
    linkcolor  = mylinkcolor!\myshade!black,
    urlcolor   = myurlcolor!\myshade!black
}

\newcommand\be{\begin{equation}}
\newcommand\ee{\end{equation}}
\newcommand\bea{\begin{eqnarray}}
\newcommand\eea{\end{eqnarray}}

\def\>#1{{\bf #1}} 

\def\k {\ell}

\renewcommand{\cosh}{\text{ch}\,}
\renewcommand{\sinh}{\text{sh}\,}

\title{\vspace{-2cm}
    \textbf{Noncommutative Lightcones from\\Quantum SO(2,1) Conformal Groups}}
\date{
    \today
    \vspace{-20pt}}
\author{
	Martina Adamo\footnote{marti.adamo@gmail.com, madamo@ubu.es (corresponding author)}~, Angel Ballesteros\footnote{angelb@ubu.es}
    ~and~Flavio Mercati\footnote{flavio.mercati@gmail.com}
	\vspace{5pt}\\
	\small Departamento de F\'isica, Universidad de Burgos, 09001 Burgos, Spain
    \vspace{10pt}}

\begin{document}

\maketitle
\begin{abstract}
\noindent Five new families of noncommutative lightcones in $2+1$ dimensions are presented as quantizations of the inequivalent Poisson homogeneous structures that emerge when the lightcone is constructed as a homogeneous space of the SO(2,1) conformal group. Each of these noncommutative lightcones maintains covariance under the action of the respective quantum deformation of the SO(2,1) conformal group. We discuss the role played by SO(2,1) automorphisms in the classification of inequivalent Poisson homogeneous lightcones, as well as the geometric aspects of this construction. The localization properties of the novel quantum lightcones are analyzed and shown to be deeply connected with the geometric features of the Poisson homogeneous spaces.
\end{abstract}

\noindent {\sc Keywords:} Noncommutative spacetimes, quantum groups, Poisson--Lie groups, homogeneous spaces, quantization, conformal symmetry, localization


\section{Introduction}
\label{sec1}

Quantum Field Theory (QFT) on noncommutative spacetimes was initially proposed in 1947 \cite{Snyder1947} as a way to cut off the ultraviolet divergences of quantum electrodynamics. The idea gained new momentum after the limitations of describing quantum gravity as a perturbative QFT were highlighted. In particular, the emergence of the Planck length as the natural scale at which quantum gravity breaks down as an effective field theory strongly suggests that there are limitations to the localizability of spacetime events \cite{Hossenfelder:2012jw}, and the standard smooth structure of spacetime might have to be replaced with a new mathematical description. After dramatic progress in the 1970s-80s on the foundations of such a description \cite{Connes1985,Woronowicz1987_1,Woronowicz1987_2}, we now have noncommutative formulations of all the basic tools of geometry and topology, which allow us, in principle, to study QFT on noncommutative spaces \cite{Vitale:2023znb}. Quantum groups \cite{majid,majid_2002,Chari} are the noncommutative generalization of Lie groups and play a central role in the description of the symmetries of noncommutative spaces \cite{Szabo:2001kg,aschieri2011noncommutative,Wulkenhaar2019}. The `semiclassical' counterpart of quantum groups are Poisson--Lie (hereafter PL) groups, which are standard Lie groups equipped with a Poisson bracket that is compatible with the group structure \cite{Chari,Drinfeld:1983ky}. A PL group contains all the essential structural information concerning the noncommutativity properties induced by the associated quantum group, with the exception of ordering ambiguities that may arise in the quantization process. The advantage of working with PL groups instead of their quantum counterparts is their computational simplicity, while still capturing most of the nonclassical aspects of a noncommutative space.

A key open question in the physics of noncommutative spaces is the fate of causality. These theories inherently possess a weakened notion of locality (see, \textit{e.g.}, \cite{Lizzi:2018qaf,Lizzi:2019wto}), which is bound to interact in unpredictable ways with the notion of causality, another pillar of modern QFT and classical relativistic theories such as General Relativity. In particular, the possibility of fuzzy/noncommutative lightcones emerging in noncommutative QFTs has been contemplated in \cite{Arzano2018,Mercati:2018ruw,Mercati:2018hlc,Franco:2023znz}. Moreover, recent works have explicitly highlighted that the difficulties of defining a sharp notion of causal relations between `points' in certain noncommutative spacetimes can be circumvented by making use of their associated noncommutative spaces of worldlines \cite{BGH2019worldlinesplb,BGH2022light,Ballesteros_2022,Ballesteros_2023}.

To be more precise, once a noncommutative Minkowski spacetime is constructed, this noncommutativity is commonly interpreted as an algebraic means of introducing a certain fuzziness generated by quantum gravity effects (for a recent review, see \cite{Addazi:2021xuf}). The spacetime position of a given particle is given by the expectation value of appropriately defined operators, and their noncommutativity leads to non-vanishing uncertainty relations among these operators. If, for instance, massless particles are considered, causality preservation requires that the induced fuzziness remains strictly confined to the lightcone. Similarly, for massive particles, their spacetime fuzziness should remain within their respective timelike or spacelike regions. However, this causality issue cannot be easily solved within the conventional approach to noncommutative Minkowski spacetimes as ambient spaces aiming to describe quantum gravity effects for all kinds of particles simultaneously, since the fuzziness around a point in the lightcone could, in principle, invade both the timelike and the spacelike regions.

In this paper, we present a novel approach to this problem by explicitly constructing noncommutative lightcones in three spacetime dimensions, seen as quantum homogeneous spaces of a quantum SO(2,1) group. In this framework, the noncommutative lightcone is constructed with no reference to any ambient noncommutative space, thus avoiding any causality issues arising from the fuzziness of the embedding. In this way, five new noncommutative lightcones with quantum conformal group invariance will be presented and analyzed.

In the next section, the classical (both future and past) lightcones ${\cal L}_\pm$ will be constructed as homogeneous spaces that are obtained by quotienting the SO(2,1) group by the stabilizing subgroup of a lightlike vector, which we will call K. Under the appropriate SO(2,1) parametrization, the orbits of a lightlike vector under the (Lorentz) group action coincide with the future- or past-oriented half lightcones. The conformal group SO(2,1) then acts as isometry group on the lightcone, and a geometric picture involving the underlying conformal symmetries emerges.

Subsequently, to define an intrinsic noncommutative structure on the lightcone, we consider its Poisson homogeneous (PH) structures. In \cref{sec3}, we introduce the PL structures on SO(2,1), which fall into three distinct families (up to Lie algebra automorphisms), characterized by three classes of classical $r$-matrices \cite{Reyman, Gomez, LTso21}. At this point, a deeper insight into SO(2,1) automorphisms turns out to be needed, since we realized that some of them do not preserve the quotient structure ${\cal L_\pm}=$SO(2,1)$/$K defining the lightcones, a fact that -- to the best of our knowledge -- has not been previously considered in the literature. The complete study of SO(2,1) automorphisms acting on the lightcones ${\cal L_\pm}$ is presented in \cref{sec4}. As a consequence, two of the families of SO(2,1) $r$-matrices split into two inequivalent classes.

Therefore, we have to consider five different families of PL structures on SO(2,1), which give rise to five non-isomorphic PH space structures on the lightcones that are explicitly presented in \cref{sec5}. This is a consequence of the property, shared by all five families, of coisotropy with respect to the subgroup K (see, \textit{e.g.}, \cite{Lu1990thesis,Ciccoli2006,BMN2017homogeneous,BGM2019coreductive}).

The associated noncommutative lightcones ${\cal L}_\pm^\k$ can then be obtained in \cref{sec6} as the quantization of the associated PH spaces, defining the algebras generating the corresponding quantum homogeneous spaces \cite{Dijkhuizen1994}. Following the approach introduced in \cite{Lizzi:2018qaf,Lizzi:2019wto,BGGM2021fuzzy}, the limitations to the localization of regions on the lightcones imposed by such noncommutative geometries are studied in terms of the representation theory of the corresponding noncommutative algebras, and they turn out to be directly connected with the geometric properties of the PH lightcones they come from. Finally, some remarks and open problems close the paper.

\section{The lightcone in 2+1 dimensions as a homogeneous space}
\label{sec2}

The Lie algebra $\mathfrak{so}(2,1)$ of the pseudo-orthogonal group in 2+1 dimensions, expressed in terms of the antisymmetric generators $\M_\gi{IJ}=-\M_\gi{JI}$, is given by
\begin{equation}\label{SO algebra}
    [\M_\gi{IJ},\M_\gi{KL}]=\eta_\gi{IL}\M_\gi{JK}+\eta_\gi{JK}\M_\gi{IL}-\eta_\gi{IK}\M_\gi{JL}-\eta_\gi{JL}\M_\gi{IK}\,,
\end{equation}
where $\eta_\gi{IJ}=\mathrm{diag}(-1,1,1)$ is the metric of the three-dimensional ambient Minkowski space, with coordinates labeled as $\text{I}, \text{J}, \text{K}, \dotsc=0,1,2$. 

The orbits of this group in the 3D ambient Minkowski space belong to three classes: spacelike two-sheeted `mass' hyperboloids, one-sheeted timelike hyperboloids, and the future and past lightcones, from which the origin $(0,0,0)$ can be singled out as a fixed point under the action of the group. In this paper, we are interested in the lightcones. Each of them can be defined as the homogeneous space obtained as the quotient of SO(2,1) by the isotropy subgroup of a lightlike vector. If the vector is future-oriented, one gets the future lightcone, and if it is past-oriented, the past lightcone is obtained.

To explicitly build the lightcones, we need to introduce the defining representation of the $\mathfrak{so}(2,1)$ Lie algebra as three-dimensional real matrices ($a,b,c,\dotsc =0,1,2$)
\begin{equation} \label{Eq:Short_representation_conformal_algebra_conf(n)}
    \rho(\M_\gi{IJ})^a{}_b = \delta_\gi{I}{}^a \, \eta_{\gi J b} -  \delta_\gi{J}{}^a \, \eta_{\gi I b} \,.
\end{equation}
Without loss of generality, we can choose the following lightlike stabilizing vectors as
\begin{equation}\label{Eq:StabilizerVector}
    v^a_\pm = \pm (l,0,-l) \,, \qquad l>0 \,,
\end{equation}
where $v^a_+$ generates the future lightcone and $v^a_-$ generates the past one. The positive quantity $l$ is a dimensional constant that gives the unit of length to the ambient coordinates. Both $v^a_+$ and $ v^a_-$ have the same isotropy subgroup, which is generated by $\K=\M_{21}-\M_{01}$.

Note that the the Lorentz group of $(d+1)$-dimensional Minkowski space and the conformal group of the $(d-1)$-dimensional Euclidean space, $\Conf(d-1)$, are isomorphic. The Lie algebras of the two groups coincide, and the natural basis of the conformal algebra $\conf(1)$, made of the generators of translations $\T$, dilations $\D$, and special conformal transformations $\K$, is related to the basis \eqref{SO algebra} of $\mathfrak{so}(2,1)$ by the following linear redefinition:
\begin{equation}\label{isomorphism Conf(n) SO(n+1 1)}
    \D=\M_{02}\,, \qquad \T=\M_{01}+\M_{21}\,,\qquad \K=\M_{21}-\M_{01}\,.
\end{equation}
The generator $\K$ is the only one that stabilizes the vector \eqref{Eq:StabilizerVector}, since
\begin{equation}
    \rho(\D)^a{}_b \, v^b_\pm = v^a_\mp \,, \qquad 
    \rho(\T)^a{}_b \, v^b_\pm = \pm 2\,l \, \delta^a{}_1 \,, \qquad 
    \rho(\K_i)^a{}_b \, v^b_\pm = 0 \,,
\end{equation}
which suggests a local factorization of the connected component to the identity of SO(2,1) (the orthochronous component $\text{SO}_+(2,1)$) that exploits this fact and reads
\begin{equation}\label{exp generators conf(1)}
    e^{\tau\,\rho(\T)}=\begin{psm}
         1+\tfrac{\tau^2}{2} & \hspace{2pt}\tau & -\tfrac{\tau^2}{2} \\
        \tau & 1 & -\tau \\
        \tfrac{\tau^2}{2} & \tau & 1-\tfrac{\tau^2}{2} 
        \end{psm},~~
    e^{\lambda\,\rho(\D)}=\begin{psm}
        \cosh \lambda \hspace{5pt}& 0 & \hspace{5pt}\sinh\lambda \vspace{5pt}\\ 
        0 & 1 & 0 \vspace{5pt}\\ 
        \sinh\lambda & 0 & \cosh\lambda 
        \end{psm},~~
    e^{\sigma\,\rho(\K)}=\begin{psm}
        1+\tfrac{\sigma^2}{2} & \hspace{2pt}-\sigma \hspace{2pt} & \tfrac{\sigma^2}{2} \\
        -\sigma & 1 & -\sigma \\
        -\tfrac{\sigma^2}{2} & \sigma & 1-\tfrac{\sigma^2}{2}
        \end{psm} \, ,
\end{equation}
where $\tau$, $\lambda$, and $\sigma$ are three parameters that, in principle, can take any value in $\mathds{R}$.
Now, a generic $\text{SO}_+(2,1)$ element is given by
\begin{equation}\label{Conf(1) element}
    \begin{aligned}
        g(\tau,\lambda,\sigma)&=e^{\tau\,\rho(\T)}e^{\lambda\,\rho(\D)}e^{\sigma\,\rho(\K)}\\
        &=\begin{psm}
            \tfrac{1}{2}(e^{\lambda}-2\tau\sigma+e^{-\lambda}(1+\tau^2)(1+\sigma^2)) & \tau-e^{-\lambda}(1+\tau^2)\sigma & \tfrac{1}{2}(e^{\lambda}-2\tau\sigma+e^{-\lambda}(1+\tau^2)(-1+\sigma^2))\\
            -\sigma+e^{-\lambda} \tau (1+\sigma^2) & 1-2e^{-\lambda} \tau\sigma & -\sigma+e^{-\lambda} \tau(-1+\sigma^2)\\
            \tfrac{1}{2}(e^{\lambda}-2\tau\sigma+e^{-\lambda}(-1+\tau^2)(1+\sigma^2)) &\hspace{5pt} \tau-e^{-\lambda}(-1+\tau^2)\sigma \hspace{5pt} & \tfrac{1}{2}(e^{\lambda}-2\tau\sigma+e^{-\lambda}(-1+\tau^2)(-1+\sigma^2))
        \end{psm}\,.
    \end{aligned}
\end{equation}
This factorization is appropriate for constructing the future and past lightcones as SO(2,1) homogeneous spaces in the form ${\cal L}_\pm = \Conf(1)/$K since by acting with a generic $\text{SO}_+(2,1)$ element \eqref{Conf(1) element} on the lightlike vector \eqref{Eq:StabilizerVector}, the action of K, factorized to the right, leaves $v^a_\pm$ invariant, thus generating a pair of two-dimensional surfaces parametrized by the group parameters $\tau$ and $\lambda$ only.

Given the local coordinates $\tau$, $\lambda$, and $\sigma$, the group axioms of SO(2,1) can be encoded into a binary multiplication law
\begin{equation}\label{composition law conf(1)}
    g(\tau_1,\lambda_1,\sigma_1)\, 
    g(\tau_2,\lambda_2,\sigma_2)
    =g\!\left( \tau_1+ \tfrac{\tau_2 e^{\lambda_1}}{1-\tau_2 \sigma_1}\,, \lambda_1+\lambda_2 - \log \! \left[ (\tau_2 \sigma_1 -1)^2 \right] \,, \sigma_2 + \tfrac{\sigma_1 e^{\lambda_2}}{1-\tau_2 \sigma_1} \right) \,,
\end{equation}
an inverse map
\begin{equation}\label{inverse conf(1)}
    g^{-1}(\tau,\lambda,\sigma) = g\!\left( \tfrac{\tau}{\tau\sigma-e^\lambda}\,, \lambda - \log\!\left[(e^\lambda-\tau\sigma)^2\right]\,, \tfrac{\sigma}{\tau\sigma-e^\lambda} \right)\,,
\end{equation}
and the specification of the coordinates of the identity element $g(0,0,0) = e$. The inverse element can be expressed in our parametrization provided that $e^\lambda - \tau\sigma\neq 0$. The multiplication map \eqref{composition law conf(1)} allows the calculation of the left- and right-invariant vector fields associated with the Lie algebra generators $\T$, $\D$, and $\K$ in the coordinates $\tau$, $\lambda$, $\sigma$
\begin{equation}\label{invariant fields conf(1)}
    \begin{aligned}
        X^L_{\T}&= e^{\lambda}\,\partial_\tau +2\sigma\,\partial_\lambda+\sigma^2\,\partial_\sigma \,,\\
        X^L_{\D}&=\partial_\lambda+\sigma \,\partial_\sigma \,,\\
        X^L_{\K}&= \partial_\sigma \,,
    \end{aligned} \qquad
    \begin{aligned}
        X^R_{\T}&=\partial_\tau \,,\\
        X^R_{\D}&=\tau\,\partial_\tau +\partial_\lambda \,,\\
        X^R_{\K}&= \tau^2\,\partial_\tau +2\tau\,\partial_\lambda+e^{\lambda}\,\partial_\sigma \,.
    \end{aligned}
\end{equation}

It is important to realize that the coordinates $(\tau,\lambda,\sigma)$ do not cover all of $\text{SO}_+(2,1)$. In fact, the spatial rotations by $\pi$ given by
\begin{equation}\label{pi rotation matrix}
e^{\pi \, \M_{12} } =     \begin{psm}
        1 \hspace{5pt}& 0 & \hspace{5pt} 0\vspace{5pt}\\ 
        0 & -1 & 0 \vspace{5pt}\\ 
        0 & 0 & -1 
        \end{psm},
\end{equation}
can only be reached through \eqref{Conf(1) element} with a singular choice of coordinates: $\tau \to \infty$,  $\sigma \to \infty$, and $\lambda \to \infty$, in such a way that $e^{-\lambda} \tau^2  \to 1$ and $e^{-\lambda} \sigma^2  \to 1$. Therefore, all transformations that involve such a $\pi$ rotation will belong to the boundary of our coordinate patch, and we can reach them by multiplying a regular group element of the form \eqref{Conf(1) element} by the matrix \eqref{pi rotation matrix}. The converse is also true: by multiplying two regular group elements, one can end up on the boundary of the coordinate patch. This feature appears in the group law \eqref{composition law conf(1)}, which becomes singular whenever $\tau_2\sigma_1=1$.

Notice also that SO(2,1) has two connected components: $\text{SO}_+(2,1)$, the orthochronous component, and  $\text{SO}_-(2,1)$, the non-orthochronous one. The three-dimensional representation of the elements of the latter can be obtained from \eqref{Conf(1) element} by multiplying $g(\tau,\lambda,\sigma)$ by the matrix
\begin{equation}\label{PTdefinition}
PT = \text{diag}(-1,1,-1)\,, \qquad \text{where} \qquad P = \text{diag}(1,1,-1)\,, \quad T = \text{diag}(-1,1,1)\,,
\end{equation}
which has unit determinant, and therefore belongs to SO(2,1), but is not connected to the identity. This matrix implements a parity transformation and a time reversal, and connects the two stabilizing vectors
\begin{equation}
    v^a_+ = (PT)^a{}_b \, v^b_- \,,
\end{equation}
so that the past lightcone can also be seen as generated by the application of non-orthochronous transformations to the future-oriented vector $v^a_+$.

The two lightcones can be described in terms of embedding coordinates $x_\pm^\gi{I}$ within the ambient (2+1)-dimensional Minkowski space. To avoid confusion between the coordinates on the group ($\tau$, $\lambda$, and $\sigma$) and those on the homogeneous space, we will replace $\tau$ with $y$ and $\lambda$ with $z$, namely
\begin{equation}\label{Eq:EmbeddingCoordinatesLightCone}
    x_\pm^\gi{I}(y,z) = g (y,z,\sigma) \, v_\pm  = \pm  e^{-z} l \,
    \begin{pmatrix}
    y^2+1\\ 2 \, y \\ y^2-1
    \end{pmatrix} \,.
\end{equation}
Notice how both coordinate $y$ and $z$ are dimensionless (they are combinations of angles and rapidities, and therefore dimensionless in natural units). It is the constant $l$ that gives the dimensions of a length (or time, which is the same in natural units) to the embedding coordinates, which satisfy the conditions
\begin{equation}\label{LightConeConditionOnEmbeddingCoords}
    \eta_\gi{IJ}  \,   x_\pm^\gi{I}(y,z)  \, x_\pm^\gi{J}(y,z)   = 0\,, \qquad \text{sgn}\!\left(x_\pm^0(y,z)\right) =  \pm 1 \,,
\end{equation}
indicating that $x_+^\gi{I}(y,z) $ defines the future lightcone ${\cal L}_+$, and $x_-^\gi{I}(y,z) $ the past one, ${\cal L}_-$. 

The action of a generic $\text{SO}_+(2,1)$ conformal group element \eqref{Conf(1) element} on ambient coordinates is
\begin{equation}
    (x_\pm')^\gi{I}(y,z) =   g(\tau,\lambda,\sigma)^\gi{I}{}_\gi{J} x^\gi{J}(y,z) =  g(\tau,\lambda,\sigma)^\gi{I}{}_\gi{J}  g (y,z,\sigma)^\gi{J}{}_\gi{K} \, v_\pm^\gi{K}    = x_\pm^\gi{I}(y',z') \,,
\end{equation}
which means that coordinates $y$ and $z$ transform according to \eqref{composition law conf(1)}, namely
\begin{equation}\label{Eq:TransformationLawHomogeneousSpaceCoords}
    y'= \tau + \tfrac{y\,e^\lambda}{1-\sigma \, y} \,, \qquad z' = z + \lambda  - \log\!\left[(\sigma \, y -1)^2\right] \,.
\end{equation}
Notice that the ray $x^0 = x^2$ lies on the boundary of the coordinate patch covered by $y$ and $z$, since to reach this light ray, we must simultaneously take the limits $y \to \infty$ and $z \to \infty$, in such a way that $e^{-z} y^2$ remains finite. In fact, the ray $x^0 = x^2$ is connected to the origin of the homogeneous space by the transformation \eqref{pi rotation matrix}.
\begin{figure}[ht!]
    \centering
    \includegraphics[width=.5\textwidth]{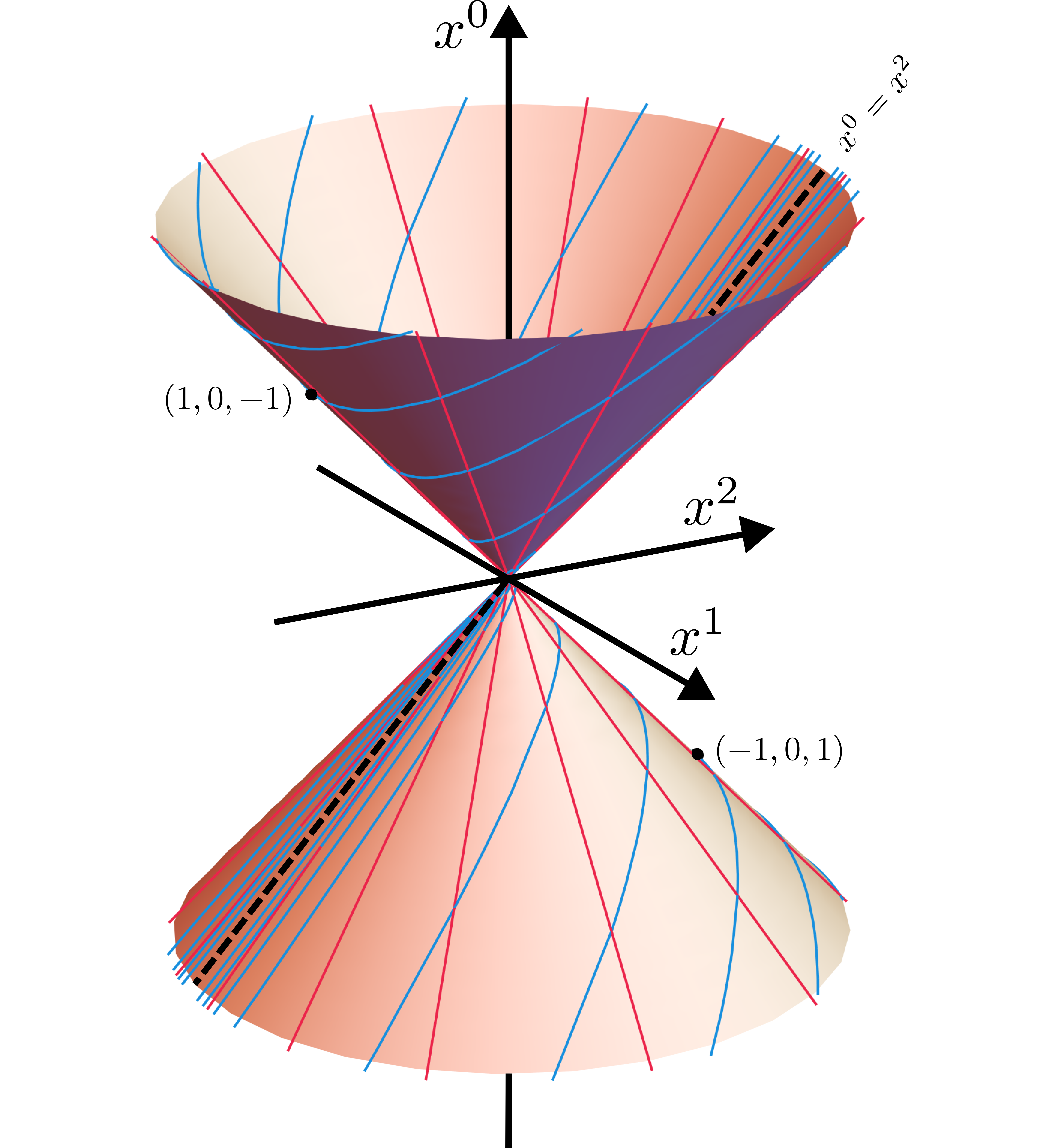}
    \captionsetup{width=.9\linewidth}
    \caption{\small The future and past lightcones in the three-dimensional Minkowski ambient space. The coordinate lines of $y$ (in blue) are obtained through pure translations and lie on the intersection of the cone with $45^\circ$ planes. Those of $z$ (in red) correspond to dilations and lie on the intersections of the cone with vertical planes (parallel to the time axis $x^0$). The black dots represent the origin $(1,0,-1)$ of the future lightcone and the origin $(-1,0,1)$ of the past lightcone when constructed as the homogeneous spaces ${\cal L}_\pm$. Notice that the boundary of our coordinate patch is the light ray $x^0=x^2$ (black dashed line).}
    \label{past and future}
\end{figure}

It is worth stressing that the geometrical interpretation of $\text{SO}(2,1) \simeq \Conf(1)$ as the group of conformal isometries of the real line is recovered by further quotienting by the dilations, which move the point on the lightcone along a light ray, either towards the origin or towards infinity (see \cref{past and future}). Quotienting the future lightcone by the action of dilations then yields the \textit{celestial sphere}, which is the space of future-directed (or past-directed) light rays, that is, the future or past boundary of the lightcone. This space can be coordinatized with the polar angle $\theta = \arctan(x^2 / x^1)$, and then mapped to the real line via a stereographic projection (excluding one point, the point at infinity, which coincides with the light ray $x^0 = x^2$ that is not covered by the coordinates $y$, $z$)
\begin{equation}
    \frac{\cos \theta}{1-\sin\theta} = \frac{x^2+\sqrt{(x^1)^2+(x^2)^2}}{x^1}  = y \,,
    \label{geoy}
\end{equation}
which provides a geometrical interpretation of the coordinate $y$. As shown in \eqref{Eq:TransformationLawHomogeneousSpaceCoords}, the action onto $y$ of translations and dilations is linear: $y'= y +\tau$, $y '= e^\lambda y$, while special conformal transformations coincide with an inversion, followed by a translation, followed by a second inversion, $y'= y / (1 - \sigma \, y)$.

\section{Poisson--Lie groups and Poisson homogeneous spaces}
\label{sec3}

As it is well known, a PL structure on a Lie group is the semiclassical limit of a quantum group: a Hopf algebra \cite{Drinfeld1987icm,Chari,majid,Abe} that represents a noncommutative deformation of the algebra of functions on a Lie group. Conversely, a quantum group is a Hopf algebra quantization of a given PL group.

The aim of this paper is to construct noncommutative analogues of the lightcones ${\cal L}_\pm$, which must be covariant under the corresponding quantum SO(2,1) conformal groups. To achieve this, we adopt the `semiclassical' approach used for constructing the noncommutative Minkowski and (A)dS spacetimes in \cite{BGH2019kappaAdS3+1,BGH2021lorentz}, as well as the noncommutative spaces of worldlines given in \cite{BGH2019worldlinesplb,BGH2022light,Ballesteros_2022,Ballesteros_2023} as quantizations of PH spaces that are covariant under the action of the corresponding PL groups. 

Therefore, we have to begin by selecting a (coboundary) PL structure $\Pi$ on the Lie group SO(2,1) with Lie algebra $\mathfrak{so}(2,1)$ given by a classical $r$-matrix defined on $\mathfrak{so}(2,1)\wedge \mathfrak{so}(2,1)$. The Poisson bivector defining $\Pi$ is explicitly provided by the Sklyanin bracket
\begin{align}\label{bb}
\{f,g\} = r^\gi{IJ} \left( X^L_\gi{I} f \, X^L_\gi{J} g - X^R_\gi{I} f \, X^R_\gi{J} g \right)\,, \qquad f,g \in \mathcal C^\infty \left(\text{SO}(2,1)\right)\,,
\end{align} 
where $X^L_\gi{I}$ and $ X^R_\gi{I}$ are the left- and right-invariant vector fields on SO(2,1) given by \eqref{invariant fields conf(1)}.

Moreover, the tangent counterpart of a coboundary PL group $(\text{SO}(2,1),\Pi)$ is a coboundary Lie bialgebra $(\mathfrak{so}(2,1), \delta)$, where the Lie bialgebra cocommutator map $\delta\,:\, \mathfrak{so}(2,1) \rightarrow  \mathfrak{so}(2,1)\wedge \mathfrak{so}(2,1)$ is obtained from the $r$-matrix through the expression
\begin{equation}\label{ba}
    \delta(\X) = [\X \otimes 1 + 1 \otimes \X,r], \qquad \forall\ \X \in \mathfrak{so}(2,1) \,.
\end{equation}

The theory of PH spaces (see \cite{Lu1990thesis,Ciccoli2006,BMN2017homogeneous,BGM2019coreductive} and references therein) states that PL groups $(\text{SO}(2,1),\Pi)$ will lead to PH structures $({\cal L}_\pm,\pi)$ on the lightcones provided that the associated Lie bialgebra $(\mathfrak{so}(2,1),\delta)$ satisfies the so-called coisotropy condition for the cocommutator $\delta$ with respect to the isotropy subalgebra of the homogeneous space, in this case generated by $\K$
\begin{equation}\label{coisotropy}
    \delta(\K) \subseteq \K \wedge \mathfrak{so}(2,1) \,.
\end{equation}
Then the Poisson structure $\pi$ on the homogeneous space is obtained by canonical projection from the corresponding $\Pi$ on SO(2,1) given by \cref{bb}. In that case, we have a PH lightcone, in which the additional Poisson bracket $\pi$ is compatible with the action of the conformal group. Moreover, when the stronger condition
\begin{equation}\label{coisotropy2}
    \delta\left(\K\right) \subseteq\K \wedge \K ,
\end{equation}
is met, $\K$ becomes a sub-Lie bialgebra, leading to a PH space of Poisson-subgroup type. In our case, since K is one-dimensional, this situation is only realized when $\delta(\K)=0$.

It is also  well known that the Lie algebra $\mathfrak{sl}(2,\mathbb R)\simeq \mathfrak{so}(2,1)$ admits three families of inequivalent Lie bialgebra structures, which correspond to three distinct families of PL structures (see \cite{Reyman, Gomez} and also \cite{LTso21}, where the explicit construction of the corresponding quantum groups and algebras is provided). Expressed in terms of the SL$(2,\mathbb{R})$ generators $\I_+$, $\I_-$, $\I_3$ with commutation relations
\begin{equation}\label{Eq:SL2_algebra}
    [\I_3,\I_\pm]=\pm 2 \, \I_\pm\,, \qquad [\I_+,\I_-]=\I_3\,,
\end{equation}
the three classical $r$-matrices generating these PL structures are
\begin{equation}\label{rmatrices sl}
    r_I=\alpha \, \I_+ \wedge \I_-\,, \qquad r_{II}=\beta\, \I_3 \wedge (\I_++\I_-)\,, \qquad r_{III}=\I_3\wedge \I_+ \,,
\end{equation}
where $\alpha$ and $\beta$ are real numbers that parametrize each family of Lie bialgebras (different values of $\alpha$ and $\beta$ correspond to inequivalent $r$-matrices, not related to each other by Lie algebra automorphisms). The $r$-matrix $r_{III}$ does not have a corresponding parameter, because its rescaling by a real number is induced by an automorphism.

In this paper, we are interested in expressing the SO(2,1) group and its $r$-matrices in terms of the conformal generators $\T$, $\D$, and $\K$. The isomorphism between $\mathfrak{sl}(2,\mathds{R})$ and $\mathfrak{so}(2,1)$
\begin{equation}\label{Eq:basic_isomorphism}
    \I_+= -\M_{01} - \M_{12} = \K \,, \qquad 
    \I_- = -\M_{01} + \M_{12} = - \T \,,\qquad \I_3=- 2 \,\M_{02} = - 2 \, \D  \,,
\end{equation}
leads to
\begin{equation}\label{Eq:r-matrices_Conf(1)}
    r_I=   \alpha \,  \T  \wedge  \K  \,, \qquad r_{II}=   2 \, \beta\,  \, \D \wedge (\T - \K) \,,\qquad r_{III}= - 2 \, \D \wedge  \K  \,.
\end{equation}
The three $r$-matrices above are representatives of three automorphism classes at the PL group level, but it will be necessary to analyze in detail the role of such group automorphisms at the level of the coisotropic PH spaces. To the best of our knowledge, this aspect has not been previously treated in the literature, and it will turn out to be essential, since in doing so we will show that more than three inequivalent PH lightcones exist and have to be considered.

First, we recall that the automorphisms of the group SO(2,1) are all inner, meaning that they can all be obtained by conjugation by an element of the group
\begin{equation}\label{Automorphisms_definition}
    g(  \tau,  \lambda,  \sigma) \to h \, g( \tau, \lambda, \sigma) \, h^{-1} \,, \qquad h \in \text{SO}(2,1) \,. 
\end{equation}
To discuss the effect of automorphisms on the PL groups and PH spaces, we need to study their action on the $r$-matrices, which in turn requires the action on the Lie algebra generators. Given a basis of linearly independent generators $\X_i \in \mathfrak{so}(2,1)$, $i=1,2,3$, the action of $h$ from \eqref{Automorphisms_definition} on each element $\X_i$ is given by
\begin{equation}\label{automorphisms on generators}
    \X_i \to h (\phi_1,\phi_2,\psi) \, \X_i \, h^{-1} (\phi_1,\phi_2,\psi) \,,
\end{equation}
where $\phi_1$, $\phi_2$, $\psi$ are three parameters on the group. The three PL groups coming from \eqref{Eq:r-matrices_Conf(1)} are invariant under automorphisms of SO(2,1), so any automorphism of the form \eqref{Automorphisms_definition} induces a recoordinatization of the group manifold that in turn induces a Poisson map between equivalent PL structures.

Nevertheless, the situation changes when PH spaces associated to certain automorphism-equivalent PL groups are considered. As we showed in the previous section, these homogeneous spaces are obtained, geometrically, by applying the group transformations to a point of the ambient Minkowski space in which the representation \eqref{Conf(1) element} of SO(2,1) lives. Only the coordinates $\tau$ and $\lambda$ act non-trivially on the lightlike point we chose as origin of the homogeneous space, and the $\sigma$ coordinate does not contribute. This implies that the automorphisms on the SO(2,1) group that just mix the $\tau$ and $\lambda$ coordinates project onto coordinate changes on the homogeneous space and can be said to preserve said homogeneous space (or to preserve the quotient structure).

Therefore, even though all group automorphisms are Poisson maps for the Poisson structure $\Pi$ on the PL group, this property does not fully extend to the PH space obtained by taking the quotient: only automorphisms that simply mix $\tau$ and $\lambda$ coordinates remain Poisson maps for the Poisson structure $\pi$ induced on the PH space. We will refer to these as `irrelevant' automorphisms, since they only correspond to coordinate changes on the PH space without changing the Poisson structure.

On the other hand, there may be some automorphisms that do not act as Poisson maps for $\pi$, despite being Poisson maps for $\Pi$ on the PL group. These `relevant' automorphisms involve mixing the $\sigma$ coordinate with $\tau$ and $\lambda$ and thus lead to genuinely new Poisson structures on the PH space. As a result, $r$-matrices connected through these relevant automorphisms generate distinct Poisson structures on the PH space, yielding more than one inequivalent PH space from a single PL group.

In our setup, we have three $r$-matrices in \eqref{Eq:r-matrices_Conf(1)}, each representing an equivalence class of automorphic $r$-matrices that induce the same Poisson structure on the group. To construct all inequivalent PH spaces, we will divide these equivalence classes into subclasses, where each subclass contains $r$-matrices connected by automorphisms that are Poisson maps for the PH space’s Poisson structure $\pi$. Consequently, in order to construct all inequivalent PH spaces, we may need to consider additional $r$-matrix representatives beyond the initial three in \cref{Eq:r-matrices_Conf(1)}, one for each equivalence subclass. This will become clearer in the next section, where we present an explicit construction of these subclasses.

\section{SO(2,1) \texorpdfstring{$\bm r$}{}-matrices and quotient-preserving automorphisms}
\label{sec4}

To make the above considerations explicit, we describe the most generic automorphism map \eqref{Automorphisms_definition} on SO(2,1) as follows: when belonging to the connected component of the identity, $h \in \text{SO}_+(2,1)$, it will be written for convenience as
\begin{equation} \label{total automorphism}
    h = e^{\phi_2 \, \K} \, e^{\phi_1 \, \D} \, e^{\psi \, \T} \,,
\end{equation}
which corresponds, through Eq.~\eqref{Automorphisms_definition}, to the following transformation rule on the group coordinates:
\begin{equation}\label{Automorphisms_expressions}
        \begin{aligned}
            \tau &\to \left( \tfrac{e^{-\phi_1}(1 + \sigma \psi)}{(\tau + \psi)(1 + \sigma \psi) - e^\lambda \, \psi} - \phi_2 \right)^{-1}\,,
            \\
            \lambda &\to \lambda - 2\log\!\left(1 + \sigma \, \psi + e^{\phi_1} \phi_2 \left( e^\lambda \psi - (\tau + \psi)(1 + \sigma \psi)\right)\right)\,,
            \\
            \sigma &\to \frac{e^{\lambda } \phi _2 \left(\psi  e^{\phi _1} \phi _2-1\right)-\left(e^{\phi _1} \phi _2 (\sigma  \psi +1)-\sigma \right) \left(\phi _2 (\tau +\psi )-e^{-\phi _1}\right)}{(\sigma  \psi +1) \left(e^{\phi _1} \phi _2 (\tau +\psi )-1\right)-\psi  \phi _2 e^{\lambda +\phi _1}} \,.
        \end{aligned}
\end{equation}
The parametrization chosen for $h$ does not cover all of $SO_+(2,1)$, because (just like the parametrization \eqref{Conf(1) element}) it excludes the rotation \eqref{pi rotation matrix}. The effect on the coordinates of conjugating $g(\tau,\lambda,\sigma)$ with the matrix \eqref{pi rotation matrix} is found to be
\begin{equation}\label{PiRotation}
    \tau \to   \frac{\sigma }{e^{\lambda }-\sigma  \tau } \,, \qquad
    \lambda \to \lambda - 2 \, \log \left(e^{\lambda }-\sigma  \tau \right) \,,\qquad
    \sigma \to \frac{\tau }{e^{\lambda }-\sigma  \tau } \,.
\end{equation}
Finally, the automorphisms generated by the non-orthochronous component $\text{SO}_-(2,1)$ can be obtained from \eqref{Automorphisms_definition} by further conjugating by the $PT$ matrix \eqref{PTdefinition}. The conjugation $ g(  \tau,  \lambda,  \sigma) \to (PT) \, g( \tau, \lambda, \sigma) \, (PT)^{-1}$ acts on the group coordinates as follows:
\begin{equation}\label{PTconjugation}
    \tau \to - \tau \,, \qquad
    \lambda \to  \lambda \,, \qquad
    \sigma \to - \sigma \,.
\end{equation}

A quick inspection of \cref{Automorphisms_expressions} reveals that the automorphisms that do not mix the $\sigma$ coordinate with $\tau$ and $\lambda$ are those such that $\psi =0$, and only those. The conjugation by the $PT$ matrix \eqref{PTconjugation} preserves the quotient as well. The only transformations that do not preserve the quotient, and therefore are relevant and can generate inequivalent PH spaces, are those of the form \eqref{Automorphisms_expressions} that involve a nonzero $\psi$ parameter, or those that involve the rotation \eqref{PiRotation} of $\pi$ of the 1 and 2 axes. Therefore, we will need only to focus on these relevant automorphisms. We may as well set the parameters $\phi_1$ and $\phi_2$ to zero, because, as can be verified, the effect of a transformation of the form \eqref{Automorphisms_expressions} on the PH space is the same as that of a transformation with the same $\psi$ and $\phi_1=\phi_2=0$, namely
\begin{equation}\label{Physical_Automorphisms_expressions}
    \tau \to  \tau + \psi \, \left(  1 - \tfrac{e^\lambda}{1 + \sigma \psi} \right)\,,\qquad
    \lambda \to \lambda - 2 \, \log\left(1 + \sigma \, \psi \right) \,,\qquad
    \sigma \to \frac{  \sigma }{1 + \sigma \,  \psi} \,,
\end{equation}
modulo changes of coordinates on the homogeneous space. In conclusion, from now on we will only focus on automorphisms generated by \eqref{pi rotation matrix} and \eqref{total automorphism} with $\phi_1=\phi_2=0$. 

The action \eqref{automorphisms on generators} of these relevant automorphisms on the generators in the conformal basis, $\X_1= \T$, $\X_2 = \D$, $\X_3 = \K$, induce the following transformation rules:
\begin{equation}\label{Rotation_algebra_automorphisms}
    \T\to\K\,, \qquad \D\to-\D\,, \qquad \K\to\T\,,
\end{equation}
and
\begin{equation}\label{Physical_algebra_automorphisms}
    \T\to\T\,, \qquad \D\to\D-\psi \T\,, \qquad \K\to \K+\psi^2\T-2\psi\D \,,
\end{equation}
respectively. In particular, the latter introduce an additional parameter in the $r$-matrices \eqref{Eq:r-matrices_Conf(1)}:
\begin{subequations}\label{r_matrix_autom_families_1}
    \begin{align}
    \begin{split}\label{r1}
         r_{I}&=\alpha(\T\wedge\K-2\psi\T\wedge\D)\equiv\alpha_1 \T\wedge(\K+\alpha_2\D) =   r_{I}(\alpha_1,\alpha_2)\,,
    \end{split}\\[0.75ex]
    \begin{split}\label{r2}
        r_{II}&=-2\beta(\D\wedge\K-\psi\T\wedge\K+(1+\psi^2)\T\wedge\D)\\
        &\equiv \beta_1((\D\wedge\beta_2 \T)\wedge\K+(1+\beta_2^2)\T\wedge\D)  =   r_{II}(\beta_1,\beta_2)\,,
    \end{split}\\[0.75ex]
    \begin{split}\label{r3}
        r_{III}&=-2(\D\wedge\K-\psi\T\wedge\K+\psi^2\T\wedge\D)\\
        &\equiv -2((\D-\gamma \T)\wedge\K+\gamma^2 \T\wedge\D)  =   r_{III}(\gamma)\,,
    \end{split}
    \end{align}
\end{subequations}
where we relabeled our parameters $\alpha_1$, $\alpha_2$, $\beta_1$, $\beta_2$ and $\gamma$ for ease of reading.

Although the transformation we used to introduce a dependence of the $r$-matrices on the parameter $\psi$ does not correspond to a recoordinatization on the lightcones, some pure recoordinatization automorphisms of the form \eqref{total automorphism} with $\psi=0$ could relate some of the $r$-matrices in \eqref{r_matrix_autom_families_1} with each other. This happens for the following subsets of $r$-matrices: in the family $r_I$~\eqref{r1}, we can connect two $r$-matrices of the family with the same $\alpha_1$ but different non-zero $\alpha_2\neq\alpha'_2$ with recoordinatization parameters:
\begin{equation}
    \phi_1=\log\tfrac{\alpha'_2}{\alpha_2}\,,\ \phi_2=0\,,\qquad\Rightarrow\qquad r_I(\alpha_1,\alpha_2)\xrightarrow[\phi_1,\phi_2]{}r_I(\alpha_1,\alpha'_2)\,.
\end{equation}
This means that all the $r$-matrices of this family with a non-zero $\alpha_2$ are actually connected by a recoordinatization, so we can simply choose a representative for them. The $r$-matrices with $\alpha_2=0$ are disconnected from the ones with $\alpha_2\neq 0$. Therefore, we split this family into two uniparametric families
\begin{equation}
    r_{I}(\alpha_1,\alpha_2=0)=\alpha_1 \T\wedge\K\,,\qquad r_{I}(\alpha_1,\alpha_2=1)=\alpha_1 \T\wedge(\D+\K)\,.
\end{equation}
The same happens for the second family $r_{II}$ \eqref{r2}:
\begin{equation}
    \phi_1=\log\tfrac{1+\beta'_2}{1+\beta_2}\,,\ \phi_2=\tfrac{-\beta_2+\beta'_2}{1+\beta'_2}\,,\qquad\Rightarrow\qquad r_{II}(\beta_1,\beta_2)\xrightarrow[\phi_1,\phi_2]{}r_{II}(\beta_1,\beta'_2)\,,
\end{equation}
with the difference that this transformation is allowed for any value of the parameters, so we have a single uniparametric family
\begin{equation}
    r_{II}(\beta_1,\beta_2=0)=\beta_1\D\wedge(\K-\T)\,.
\end{equation}
Finally, in the case of the third family $r_{III}$ \eqref{r3}, for non-zero values of the parameter $\gamma\neq\gamma'$, we can connect two $r$-matrices of the family with the following recoordinatization parameters:
\begin{equation}
    \phi_1=2\log\tfrac{\gamma'}{\gamma}\,,\ \phi_2=\tfrac{\gamma-\gamma'}{\gamma'^2}\,,\qquad\Rightarrow\qquad r_{III}(\gamma)\xrightarrow[\phi_1,\phi_2]{}r_{III}(\gamma')\,.
\end{equation}
This shows that all non-zero values of the parameter $\gamma$ are connected by a recoordinatization. Therefore, we again have a splitting into two subcases:
\begin{equation}
    r_{III}(\gamma=0)=-2 \, \D\wedge\K\,,\qquad r_{III}(\gamma=1) = 2 \, (\T - \D) \wedge\K - 2\, \T\wedge\D \,.
\end{equation}

Thus, the five classes of inequivalent structures obtained so far are the following:
\begin{subequations}\label{five families r-matrices}
    \begin{align}
        r_{I}^A(\alpha_A)& =\alpha_A \T\wedge\K\,,\label{r1A}\\
        r_{I}^B(\alpha_B) &=\alpha_B \T\wedge(\D+\K)\,,\label{r1B}\\
        r_{II}(\beta) &=\beta \, \D\wedge(\K-\T)\,,\label{r22}\\
        r_{III}^A& = -2 \, \D\wedge\K\,,\label{r3A}\\
        r_{III}^B &= 2 \,(\T -\D )\wedge\K- 2 \, \T\wedge\D \,, \label{r3B}
    \end{align}
\end{subequations}
where $r_{I}^B$ and $r_{III}^B$ arise as new ones. We have three uniparametric families of $r$-matrices depending on the real parameters $\alpha_A,\alpha_B,\beta\in\mathbb{R}$, and two additional isolated $r$-matrices.
    
Finally, the effect of the $\pi$ rotation automorphism \eqref{Rotation_algebra_automorphisms} on the families \eqref{five families r-matrices} leads to
    \begin{equation}\label{pi rotation five families r-matrices}
    \begin{aligned}
        {r'}_{I}^A(\alpha_A) &= - \alpha_A \T\wedge\K = r_{I}^A(-\alpha_A)\,,\\
        {r'}_{I}^B(\alpha_B) &=\alpha_B \K\wedge(\T-\D)\,,\\
        {r'}_{II}(\beta) &= \beta \, \D\wedge(\K-\T) = r_{II}(\beta)\,,\\
        {r'}_{III}^A &= 2 \, \D\wedge\T\,,\\
        {r'}_{III}^B &= -2 \,(\K + \D )\wedge \T + 2 \, \K\wedge\D \,.
    \end{aligned}
    \end{equation}
The families $r_I^A$ and $r_{II}$ did not change (the automorphism just sends the parameters $\alpha_A$ and $\beta$ to $-\alpha_A$ and $\beta$), while the other three families are modified. However, upon a closer inspection, these changes correspond to the following recoordinatizations of the families we already have:
\begin{equation}
    \begin{array}{rlcc}
        \forall \ \phi_1\in\mathbb{R}\,, & \phi_2=-\tfrac{1}{2}\,, &\qquad \Rightarrow \qquad& \qquad r_I^A(-\alpha_B)\xrightarrow[\phi_1,\phi_2]{}{r'}_{I}^B(\alpha_B)\,,\\            
    \phi_1=0\,, &\phi_2=1\,, & \qquad \Rightarrow \qquad &\qquad r_{III}^B \xrightarrow[\phi_1,\phi_2]{}{r'}_{III}^A\,,\\        
    \phi_1=0\,, & \phi_2=2\,, &\qquad \Rightarrow \qquad & \qquad r_{III}^B\xrightarrow[\phi_1,\phi_2]{}{r'}_{III}^B\,.
    \end{array}
\end{equation}
Therefore, \cref{five families r-matrices} provides the full set of families of $r$-matrices to be considered in order to construct noncommutative lightcones with quantum SO(2,1) invariance.

\section{Poisson homogeneous structures on the lightcone}
\label{sec5}

The five families of $r$-matrices \eqref{five families r-matrices} correspond to the following five cocommutators that define, through \eqref{ba}, the associated Lie bialgebra structures on the $\mathfrak{so}(2,1)$ algebra:
    \begin{equation}\label{LBialgebras}
    \begin{aligned}
        r_I^A&: & &\delta(\T)=-2\alpha_A \T\wedge\D\,,  &  
        &\delta(\D)=0\,,  & &\delta(\K)=2\alpha_A \D\wedge\K\,,\\
        r_I^B&: & &\delta(\T)=-2\alpha_B \T\wedge\D\,, & &\delta(\D)=\alpha_B \T\wedge\D\,, &   &\delta(\K)=\alpha_B (2\D+\T)\wedge\K\,,\\
        r_{II}&: & &\delta(\T)=-2\beta \T\wedge\K\,, & &\delta(\D)=-\beta \D\wedge(\K+\T)\,, & &\delta(\K)=\beta \T\wedge\K\,,\\
        r_{III}^A&: & &\delta(\T)=2\T\wedge\K\,,& &\delta(\D)=2 \D\wedge\K\,,& &\delta(\K)=0\,,\\
        r_{III}^B&: & &\delta(\T)= 2 \T\wedge(\K-2\D) \,,& &\delta(\D)= 2\D \wedge (\K+\T) \,,  & &\delta(\K)= 2(2\D-\T)\wedge\K \,.
    \end{aligned}
    \end{equation}
The coisotropy condition \eqref{coisotropy} holds for all five cases above, while only the matrix $r_{III}^A$ \eqref{r3A} satisfies the Poisson subgroup condition \eqref{coisotropy2},  which makes $\K$ into a sub-Lie bialgebra.

Now, by using the Sklyanin bracket \eqref{bb}, we get five PL structures $\Pi$ on SO(2,1)
\begin{equation}\label{PLgroups}
    \begin{aligned}
        r_I^A&: &  &\{\tau,\lambda\}=-2\alpha_A\tau\,,&  &\{\sigma,\tau\}=0\,, &  &\{\sigma,\lambda\}=-2\alpha_A \sigma\,, \\
        r_I^B&: &  &\{\tau,\lambda\}=\alpha_B(e^\lambda-2\tau-1) \,,&   &\{\sigma,\tau\}=-\alpha_B e^\lambda\sigma \,, &   &\{\sigma,\lambda\}=-\alpha_B \sigma (2+\sigma)\,, \\
        r_{II}&: &  &\{\tau,\lambda\}=\beta(e^\lambda-\tau^2-1) \,,&   &\{\sigma,\tau\}=-\beta e^\lambda(\sigma-\tau) \,, &   &\{\sigma,\lambda\}=\beta(e^\lambda-\sigma^2-1) \,, \\
        r_{III}^A&: &  &\{\tau,\lambda\}=2\tau^2\,,&   &\{\sigma,\tau\}=-2e^\lambda\tau\,, &   &\{\sigma,\lambda\}=2(1-e^\lambda)\,, \\
        r_{III}^B&: &  &\{\tau,\lambda\}=-2 (e^\lambda-(\tau-1)^2) \,,&   &\{\sigma,\tau\}=2e^\lambda(\sigma-\tau)\,, &   &\{\sigma,\lambda\}=-2 (e^\lambda-(\sigma-1)^2)\,. \\
    \end{aligned}
\end{equation}
Notice that, at this PL group level, the inequivalent  structures are just three (or rather two uniparametric families plus a single, isolated case): $r_{I}^{A,B}$, $r_{II}$, and $r_{III}^{A,B}$, since the `$A$' and `$B$' subfamilies of families $I$ and $III$ are related by automorphisms. Casimir functions for all these Poisson structures \eqref{PLgroups} have been determined and are as follows:

\begin{equation}
\label{5phlc}
\begin{gathered}
C_I^A=\sfrac{\tau}{\sigma}\,,\qquad \ 
C_I^B=\sfrac{1-e^\lambda+\sigma\tau+2\tau}{2\sigma}\,,\qquad \
C_{II}=\sfrac{1-e^\lambda+\sigma\tau}{\sigma-\tau}\,,\\
C_{III}^A=\sfrac{1-e^\lambda+\sigma\tau}{\tau}\,,\qquad \ 
C_{III}^B=\sfrac{1-e^\lambda+\sigma\tau-\sigma-\tau}{\sigma-\tau}\,, 
\end{gathered}    
\end{equation}
The level sets of these Casimir functions identify the symplectic leaves of the Poisson structure.

Now, the Poisson brackets $\pi$ of the PH spaces of each of the five families \eqref{five families r-matrices} can be simply obtained by projecting the PL  brackets \eqref{PLgroups} to the first two coordinates, $\tau$ and $\lambda$, and replacing $\tau$ with $y$ and $\lambda$ with $z$
\begin{subequations}\label{HSPB}
    \begin{align}
        r_I^A&: \hspace{-0.01em}\quad \{y, z\}=-2 \, \alpha_A\, y\,, \\
        r_I^B&: \hspace{-0.01em}\quad\{ y, z\}=\alpha_B\, (e^z-2\,  y-1) \,,\\
        r_{II}&: \hspace{-0.01em}\quad\{ y, z\}=\beta\, (e^z- y^2-1) \,,\label{pbr2} \\
        r_{III}^A&: \hspace{-0.01em}\quad\{ y, z\}=2 \, y^2\,, \\
        r_{III}^B&: \hspace{-0.01em}\quad\{ y, z\}=-2 \, \left( e^z-( y-1)^2 \right) \,.
    \end{align}
\end{subequations}
We have found a set of Darboux coordinates $\{ q,p\} =1$ for each case
\begin{equation}\label{DarbouxCoordinates}
    \begin{aligned}
        r_I^A&: & & z = p  \,, & & y = \pm e^{- 2 \, \alpha_A \, (q-q_0)} \,, \\
        r_I^B&: & & z = p  \,, & & y =  \sfrac{1}{2} \left(e^p-1\right)+e^{-2  \,\alpha_B (q - q_0)}\,, \\
        r_{II}&: & & z = p  \,, & & y = -\sqrt{1-e^p} \, \tan\! \left(\sqrt{1-e^p} \, \beta \, (q-q_0)\right) \,, \\
        r_{III}^A&: & & z = p  \,, & & y =  -\left( 2 \, (q - q_0) \right)^{-1} \,, \\
        r_{III}^B&: & & z = p  \,, & & y =  1-e^{p/2} \,\tanh \!\left( 2 \,e^{p/2} \, (q-q_0)\right) \,,\\
    \end{aligned}
    \end{equation}
where $q_0$ is an integration constant and the $\pm$ symbol refers to two possible choices of Darboux coordinates.
    
The coordinates $y$ and $z$ appearing in \eqref{HSPB} are equally valid on the future and the past lightcone, $\mathcal{L}_+$ and $\mathcal{L}_-$, depending on which of the two embeddings \eqref{Eq:EmbeddingCoordinatesLightCone} is used. The Poisson brackets \eqref{HSPB} can be expressed in terms of the ambient coordinates as follows:
\begin{equation}\label{EmbeddingCoordsPB}
\{x^\gi{I}_\pm , x^\gi{J}_\pm \} = \left( u_\gi{L} \, x^\gi{L}_\pm + w_\pm \right) \, \varepsilon^\gi{IJ}{}_\gi{K} \, x^\gi{K}_\pm \,,
\end{equation}
provided that the lightcone constraint \eqref{LightConeConditionOnEmbeddingCoords} holds. The values for the parameters $u_\gi{I}$ and $w_\pm$ differ among the five families:
\begin{equation}
		\begin{aligned}
			r_I^A&: & & u_\gi{I}  =   2   \, \alpha_A \, \delta^1_\gi{I}   \,,& &  w_\pm = 0 \,, \\
			r_I^B&: & & u_\gi{I} =  \alpha_B \left(\delta^0_\gi{I}  + 2 \, \delta^1_\gi{I}  - \delta^2_\gi{I}  \right) \,, & &  w_\pm =  \mp 2 \, l \,  \alpha_B \,, \\
			r_{II}&: & & u_\gi{I} = 2   \, \beta \, \delta^0_\gi{I}  \,, & &  w_\pm  = \mp 2 \, l \,\beta       \,, \\
			r_{III}^A&: & & u_\gi{I}  = -2  \, \left(\delta^0_\gi{I} +\delta^2_\gi{I} \right)  \,, & &  w_\pm  = 0\,, \\
			r_{III}^B&: & & u_\gi{I} = 4  \, \left(\delta^1_\gi{I} - \delta^0_\gi{I}  \right)  \,, & &  w_\pm  =  \pm 4 \, l \,, \\
		\end{aligned}
	\end{equation}
where $l$ is the positive constant chosen for the stabilizing vectors \eqref{Eq:StabilizerVector}.

Notice how all the right-hand sides of Eqs.~\eqref{EmbeddingCoordsPB} are proportional to $\varepsilon^\gi{IJ}{}_\gi{K} \, x^\gi{K}_\pm $, where $\varepsilon^\gi{IJ}{}_\gi{K} $ is the 3D Levi-Civita tensor with one index lowered with the 2+1-dimensional Minkowski metric. Under inner automorphisms, the embedding coordinates transform like contravariant vectors under classical Lorentz transformations. Thus, the effect of an automorphism on the commutation relations \eqref{EmbeddingCoordsPB} is to Lorentz-transform the constants $u_\gi{I}$ as if they formed a covariant vector, while the constants $w_\pm$ are left invariant. Therefore, two models whose $u_\gi{I}$ are related by a Lorentz transformation are equivalent, in the sense that they are related by a coordinate change, as long as their $w_\pm$ values coincide.

The constant $u_\gi{I}$ is spacelike in both cases of family $I$, timelike in family $II$, and lightlike in both cases of family $III$. Consequently, apart from a rescaling, one can connect cases $r_{I}^A$ and $r_{I}^B$, as well as $r_{III}^A$ and $r_{III}^B$, using an automorphism. However, the `A' cases have $w_\pm =0$, while the `B' cases do not. The conclusion is that the `A' cases are limiting cases of the `B' ones, where $l \to 0$. Given that $l$ is the length scale of the ambient Minkowski space, sending $l$ to zero corresponds to observing the model from  infinitely far away, where the features relevant in a neighbourhood of the origin of size $\sim l$ cannot be appreciated.
\begin{figure}[ht!]
    \centering
      \makebox[\textwidth][c]{\includegraphics[width=1.03\textwidth]{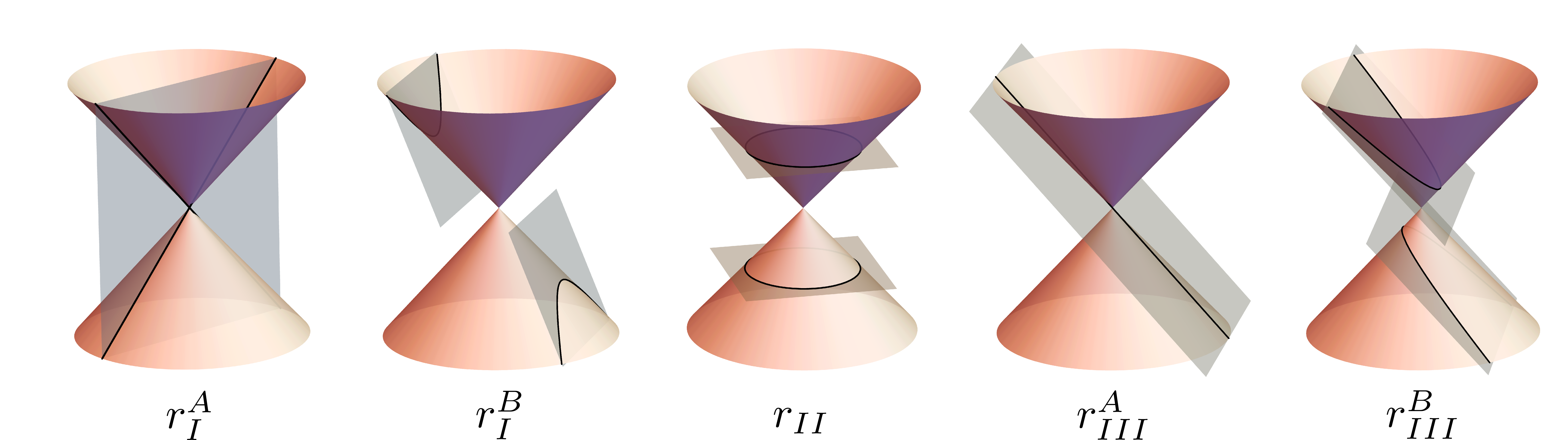}}%
    \captionsetup{width=.9\linewidth}
    \caption{\small The regions of the lightcones where the five Poisson brackets vanish are given by the black curves defined by the intersection of the lightcones with the planes shown in this figure.}
    \label{Localization plot}
\end{figure}

This rewriting makes it easier to identify the submanifolds of the lightcones where the Poisson brackets vanish in each model, which will be relevant in the quantum case concerning localization properties. These are all one-dimensional submanifolds obtained by intersecting the lightcones \eqref{LightConeConditionOnEmbeddingCoords} with the  planes $u_\gi{L} \, x^\gi{L}_\pm + w_\pm =0$, that is
\begin{equation}
    \begin{aligned}
        r_I^A&: &\quad &  x^\gi{1}_\pm = 0\,, & \qquad & \text{timelike plane through the origin,} \\
        r_I^B&: &\quad & x^\gi{0}_\pm + 2 \, x^\gi{1}_\pm - x^\gi{2}_\pm \mp 2 \, l = 0 \,,& & \text{timelike planes separated by $2\,l$,} \\
        r_{II}&: &\quad & x^\gi{0}_\pm \mp l  = 0\,,& & \text{spacelike  planes separated by $2\,l$,} \\
        r_{III}^A&: &\quad & x^\gi{0}_\pm + x^\gi{2}_\pm  = 0\,,& & \text{lightlike plane through the origin,}\\
        r_{III}^B&: &\quad & x^\gi{1}_\pm - x^\gi{0}_\pm \pm l  = 0\,,& & \text{lightlike planes separated by $2\,l$,}
    \end{aligned}
    \end{equation}
where the $\pm$ symbol refers to the future and past lightcones. We plotted these intersecting planes, for each of the five models, in \cref{Localization plot}. Notice how the parameter $l$ appearing in three of the expressions above simply sets the scale for the whole picture without changing anything substantial: for $r_I^B$, $r_{II}$, and $r_{III}^B$, it determines the vertical distance between the origin and the two intersecting planes. Independently of the value of $l$, the intersection remain consistent: two light rays in the case of $r_I^A$, two hyperbolas for $r_I^B$, two circles for $r_{II}$, one light ray for $r_{III}^A$, and two parabolas for $r_{III}^B$. 

Cases $r_I^A$ and $r_{III}^A$ can be seen as degenerate limits of $r_I^B$ and $r_{III}^B$, respectively, as $l \to 0$. This further clarifies the sense in which the `A' and `B' subcases are not mere recoordinatizations of each other.  Instead, the `A' cases are just the `B' cases viewed from infinitely far away, making the failure of the two planes to meet at the origin not appreciable (and , in the $r_I^B \to r_I^A$ case, further subject to a Lorentz transform that makes the intersecting planes vertical). Similarly, the Poisson brackets between embedding coordinates \eqref{EmbeddingCoordsPB} show that the two `A' models can be obtained from the `B' models via an automorphism of the $x^\gi{I}_\pm$ followed by the $l \to 0$ limit. In other words, the `A' models are essentially the `B' models seen from afar (and from a Lorentz-transformed perspective).

\section{Representations of noncommutative lightcones and their localizability properties}
\label{sec6}

Introducing a fuzziness parameter $\k >0$, which plays a similar role as $\hbar$ in ordinary Quantum Mechanics (see \cite{BGGM2021fuzzy} for a more detailed discussion), and promoting the coordinate functions $y$, $z$ to operators $\hat{y}$, $\hat{z}$, we can quantize \cref{HSPB}:
\begin{subequations}
\label{QuantumCommutators1}
\begin{align}
    r_I^A&: \hspace{-0.01em}\quad [\hat{y}, \hat{z}] = -2i \, \k\, \alpha_A \hat{y}\,, \label{rq1A}\\
    r_I^B&: \hspace{-0.01em}\quad [\hat{y}, \hat{z}] = i\, \k\, \alpha_B\, (e^{\hat{z}}-2 \, \hat{y}-1) \,,\label{rq1B}\\
    r_{II}&: \hspace{-0.01em}\quad [\hat{y}, \hat{z}] = i\, \k\, \beta\, (e^{\hat{z}}- \hat{y}^2-1) \,, \label{rq2}\\
    r_{III}^A&: \hspace{-0.01em}\quad [\hat{y}, \hat{z}] = 2i\, \k\,  \hat{y}^2\,, \label{rq3A}\\
    r_{III}^B&: \hspace{-0.01em}\quad [\hat{y}, \hat{z}] = -2i\, \k \left( e^{\hat{z}}-( \hat{y}-1)^2 \right) \, , \label{rq3B}
\end{align}
\end{subequations}
with no ordering issues arising. This $\k$-quantization should be understood in the sense that its semiclassical limit 
\begin{equation}
    \{\, \cdot \, , \, \cdot \,  \} = \lim_{\k \to 0} \frac{[\, \cdot \, , \, \cdot \, ]}{i \, \k} \,,
\end{equation}
recovers the five PH lightcones \eqref{HSPB}. The commutation relations \eqref{QuantumCommutators1} are supplemented by the Hermiticity conditions on the `quantum coordinates' $\hat{y}$ and $\hat{z}$
\begin{equation}
    \hat{y}^\dagger = \hat{y} \,, \qquad \hat{z}^\dagger = \hat{z} \,,
\end{equation}
which are compatible with the commutators, as can be easily verified. Notice how the fuzziness parameter $\k$ is dimensionless; it determines the magnitude of fuzziness implied by the commutation relations of the dimensionless coordinates $\hat y$ and $\hat z$. 

The irreducible regular representations of an algebra isomorphic to case $r_I^A$ in \cref{rq1A} have been described in \cite{DabrowkiPiacitelli} and further analyzed in \cite{Lizzi:2018qaf,Lizzi:2019wto}. The representation is based on the Darboux coordinates \eqref{DarbouxCoordinates}, and the operators take the form
\begin{equation}
    \hat z = - i \, \k \, \partial_q \,, \qquad \hat y = 0,\,\pm e^{-2 \, \alpha_A \, (q-q_0)} \,,
\end{equation}
where the three cases, in which $\hat y$ is respectively positive, negative, or zero, are the three Hermitian irreducible representations, and the integration constant $q_0$ just shifts the origin of the representation horizontally. The Hilbert space is that of square-integrable functions $L^2(\mathds{R})$ of $q \in \mathds{R}$.

A similar analysis applies also to the algebras in \cref{rq1B} and \cref{rq3A}: 
\begin{equation}
\begin{aligned}
    r_I^B&: & & \hat z = - i \, \k \, \partial_q  \,, & & \hat y =  \sfrac{1}{2} \left(e^{-i \, \k \, \partial_q} -1\right)+e^{-2  \,\alpha_B (q - q_0)}\,, \\
    r_{III}^A&: & & \hat z = -i \, \k \, \partial_q  \,, & & \hat y =  -\left( 2 \, (q - q_0) \right)^{-1} \,.
\end{aligned}
\end{equation}
In the case of $r_{III}^A$, the singularity at $q=0$ forces us to change the Hilbert space for a space of functions that vanish at the origin. For model $r_{III}^B$ in \cref{rq3B}, we modify the representation used for algebra $r_{III}^A$:
\begin{equation}
    \hat z = -\log\! \left(4 (q-q_0)^2 + e^{p_0 + i \, \k \, \partial_q }\right) \,, \qquad \hat y = 1 -\left( 2 \, (q - q_0) \right)^{-1} \,,
\end{equation}
with a Hilbert space consisting again of square-integrable functions that vanish at the origin.

The model $r_{II}$ in \cref{rq2} is special because one of the variables of the canonical pair we find has a compact domain, indicating that the canonically conjugate operator should have a discrete spectrum. This compactness is related to the fact, observed above, that in this model the region where the Poisson brackets vanish is a compact circle. Rather than using the Darboux variables found in \eqref{DarbouxCoordinates}, which depend on products of $q$ and $p$ and make quantization difficult, it is convenient to consider the cylindrical coordinates $\theta$, $t$ in the ambient space (already used in \eqref{geoy} for the geometrical interpretation of $y$):
\begin{equation}
    y =\frac{\cos \theta}{1-\sin\theta} \,, \qquad z = -\log\! \left(\frac{\pm t(1- \sin \theta )}{2}\right) \,,
\end{equation}
where the $\pm$ symbol refers to the future and past lightcones. In these coordinates, the Poisson brackets \eqref{pbr2} of case $r_{II}$ turn into $\{ \theta , t \} = 2\, \beta \, (t \mp 1)$, and a set of Darboux coordinates can be found as $\theta = q$, $t = \pm (e^{2 \, \beta \, (p-p_0)}+1)$, which imply:
\begin{equation}
    y =\frac{\cos q}{1-\sin q} \,, \qquad z = -  \log \left(  1- \sin q \right) - \log\!\left( \sfrac{ e^{2 \, \beta \, (p-p_0)}+1 }{2}\right) \,.
\end{equation}
We then found a choice of Darboux coordinates that is suitable for representing case $r_{II}$ of \cref{rq2}, because there is no ordering ambiguity in the map that connects them to $\hat y$, $\hat z$:
\begin{equation}\label{operators case r2}
    \hat y = \frac{\cos \theta}{1-\sin  \theta} \,, \qquad \hat z = -  \log \!\left(  1- \sin  \theta \right) -   \log \!\left( \sfrac{e^{2 \, \beta \, (\hat{L}_\theta -p_0)}+1  }{2} \right) \,.
\end{equation}
The Hilbert space will be that of square-integrable periodic functions of $ \theta \in [0,2\pi)$, and  the operator $\hat{L}_\theta$ in the expression \eqref{operators case r2} for $\hat z$ is the conjugate momentum of the angle $\theta$, which has a discrete spectrum and can be obtained as a self-adjoint extension of the derivative operator $ - i \, \k \partial_\theta$.

We now turn to the problem of localizability, \textit{i.e.}, how well points can be resolved in noncommutative spaces such as \eqref{QuantumCommutators1}. This problem can be discussed in terms of the states on the algebras \eqref{QuantumCommutators1}. These states are positive linear functionals of norm one on the algebra, and maximally-localized states can be described as limits of well-normalized states. The maximally-localized states determine how well a region can be resolved. For all models in \cref{QuantumCommutators1} except $r_{II}$ \eqref{rq2}, we can discuss the perfectly-localized states in the Darboux variables, which close the Heisenberg algebra $[\hat{q},\hat{p}] =i \k$ and have a real spectrum. The coordinates $\hat{q}$ or the momenta $\hat{p}$ can be arbitrarily well localized around a value if we consider normalized states that are increasingly more peaked around a particular value. In the limit of perfect localization, these turn into delta functions, which are not normalizable and therefore called improper states, but admit the physical interpretation of an idealization, as one can push the localization limit arbitrarily far while always maintaining a normalizable state. Thus, $\hat{p}$ or $\hat{q}$ can be separately localized arbitrarily well, but there are limits on their simultaneous localization, in accordance with the uncertainty relation
\begin{equation}
    \delta \hat{q} \, \delta \hat{p}  \geq \frac{\k}{2} \,.
\end{equation}
The optimally-localized states in both $\hat{p}$ and $\hat{q}$ are the coherent states, whose wavefunctions are Gaussians given by
\begin{equation}
    \psi(q) = {\tfrac {1}{{\sqrt {\sigma  \sqrt{2\pi} }}}}e^{- \left({\frac {q-q_0 }{2\sigma }}\right)^2 + i  \frac{p_0 \, q_0}{\k}}  ~~ \Rightarrow ~~ \langle \hat{q} \rangle = q_0 \,, ~~ \langle \hat{p} \rangle = p_0 \,, ~~ \delta \hat{q} = \sigma \,, ~~   \delta \hat{p} = \frac{\k}{2 \sigma } \,.
\end{equation}

The commutation relations of an algebra impose that, on any state, the uncertainty product is larger than the expectation value of the right-hand side. In the case of the Heisenberg algebra, the right-hand side is proportional to the identity operator so its expectation value is a constant, and one can never perfectly localize a state in phase space. When the the commutators are non-constant, as in \eqref{QuantumCommutators1}, the situation might be different. The uncertainty relations implied by our commutators are
\begin{equation}\label{UncertaintyRelations}
    \begin{aligned}
        r_I^A&: &\quad &\delta \hat{y} \,  \delta  \hat{z} \geq -  \k\,\left|  \alpha_A \, \left\langle \hat{y} \right\rangle  \right| \,, \\
        r_I^B&: &\quad &\delta \hat{y} \,  \delta  \hat{z} \geq \sfrac{\k}{2} \, \left| \alpha_B\, \left \langle e^{\hat{z}}-2 \, \hat{y}-1 \right\rangle  \right| \,,\\
        r_{II}&: &\quad &\delta \hat{y} \,  \delta  \hat{z} \geq  \sfrac{\k}{2}\, \left|\beta \, \left\langle e^{\hat{z}}- \hat{y}^2-1  \right\rangle \right|  \,, \\
        r_{III}^A&: &\quad &\delta \hat{y} \,  \delta  \hat{z} \geq \sfrac{\k}{2} \,  \left| \left\langle \hat{y}^2 \right\rangle  \right| \,, \\
        r_{III}^B&: &\quad &\delta \hat{y} \,  \delta  \hat{z} \geq  \k\, \left| \left\langle e^{\hat{z}}-( \hat{y}-1)^2 \right\rangle  \right| \,, \\
    \end{aligned}
    \end{equation}
and in all five cases there are (limits of) well-normalized states that annihilate the expectation values on the right-hand sides above.

As a first exploration of the locality structure of our noncommutative spacetimes, we can identify the regions of the lightcone where the right-hand sides of \eqref{UncertaintyRelations} vanish. Improper states defined as delta functions localized on a point of these regions will be perfectly localized, as both the uncertainties $\delta \hat z$ and $\delta \hat y$ will vanish on those states, and they will be obtainable as limits of sequences of well-normalized states on the corresponding algebras. These regions have been discussed already in \cref{sec5}, and are depicted in \cref{Localization plot} as intersections between the lightcones and certain planes.

It is worth stressing that in the special case $r_{III}^A$, in which the subgroup K is a PL subgroup, its associated noncommutative spacetime \eqref{rq3A} is just quadratic in $\hat y^2$, while all the other commutators always possess a linear term when a formal power series expansion is considered. After quantization, this makes the noncommutativity of this special case milder, because one needs to localize a state further from the line $y=0$ in order to see the fuzziness, compared to the other models.

\section{Final remarks}
\label{sec7}

To conclude, we stress that the approach here presented can be used with no modification for the case of noncommutative 3+1-dimensional lightcones, which can be constructed as quantizations of coisotropic PH structures coming from the PL structures of the SO(3,1) Lorentz group when interpreted in a conformal setting. We believe that the new noncommutative lightcones \eqref{QuantumCommutators1} presented here offer a novel twist on an important issue in noncommutative physics: the interplay between the relaxed notion of locality implied by noncommutativity and the notion of causality, which is fundamental for QFT. Indeed, a number of studies in the past have focused on the possibility of fuzzy lightcones emerging from the spacetime correlations of noncommutative QFTs \cite{Arzano2018,Mercati:2018ruw,Mercati:2018hlc,Franco:2023znz}; however, the noncommutativity we describe here stays on the lightcone. The lightcone itself is sharp, but there are limitations to localizing points \textit{on the lightcone}. This feature might be interesting from a phenomenological point of view, in light of studies of propagation of massless particles in a quantum spacetime (see \cite{Addazi:2021xuf,AlvesBatista:2023wqm}). This approach complements nicely the studies of noncommutative spaces of worldlines initiated recently \cite{BGH2019worldlinesplb,BGH2022light,Ballesteros_2022,Ballesteros_2023}, and it would be particularly interesting to compare the results of the present paper with those focusing on the spaces of noncommutative lightlike worldlines \cite{BGH2022light}. Work on all these lines is in progress and will be presented elsewhere.

\section*{Acknowledgements}

M.A., A.B., and F.M.~would like to acknowledge support from the grant PID2023-148373NB-I00 funded by MCIN/AEI/10.13039/501100011033/FEDER -- UE, the Q-CAYLE Project funded by the Regional Government of Castilla y León (Junta de Castilla y León) and the Ministry of Science and Innovation MICIN through NextGenerationEU (PRTR C17.I1).
F.M.~acknowledges support from the Agencia Estatal de Investigación (Spain) under grant CNS2023-143760.

\small

\addcontentsline{toc}{section}{References}\bibliographystyle{utphys}
\bibliography{bib}

\end{document}